\documentclass[aps,prd,twocolumn,preprintnumbers,nofootinbib,superscriptaddress,amsmath]{revtex4-2}
\usepackage[english]{babel}
\usepackage{bm}
\usepackage[utf8]{inputenc}
\usepackage[colorinlistoftodos, color=green!40, prependcaption]{todonotes}
\usepackage{amssymb}
\usepackage{graphicx}
\usepackage{natbib}

\usepackage[pdftex, pdftitle={Article}, pdfauthor={Author}]{hyperref} 

\begin{document}
\newcommand{\g}[1]{{\color{red}[G: #1]}}
\newcommand{\rs}[1]{{\color{blue}[RS: #1]}}
\title{Thawing quintessence and transient cosmic acceleration in light of DESI}

\author{Rayff de Souza}
\email{rayffsouza@on.br}
\author{Gabriel Rodrigues}
\email{gabrielrodrigues@on.br}
\author{Jailson Alcaniz}
\email{alcaniz@on.br}
\affiliation{Observatório Nacional, Rio de Janeiro - RJ, 20921-400, Brasil}

\begin{abstract}
Recent analysis of the DESI Collaboration challenges the $\Lambda$-Cold Dark Matter ($\Lambda$CDM) model, suggesting evidence for a dynamic dark energy. These results are obtained in the context of generic parameterizations of the dark energy equation of state (EoS), which better fit the data when they exhibit an unphysical phantom behavior in the past. In this paper, we briefly analyze how ambiguous this latter conclusion  can be in light of the background degeneracy between EoS parameterizations and minimally coupled
quintessence scenarios. We then investigate whether the current observational data can be accommodated with a non-phantom, thawing dark energy EoS, typical of a broad class of quintessence models. We show that the thawing behavior of this EoS performs comparabily to the Chevallier-Polarski-Linder parameterization and is statistically competitive with $\Lambda$CDM while predicting cosmic acceleration as a transient phenomenon. Such a dynamic behavior aligns with theoretical arguments from string theory and offers a way out of the trans-Planckian problem that challenges the ever-accelerated $\Lambda$CDM paradigm.

\end{abstract}



\maketitle


\section{Introduction}\label{sec:1}


Measurements of Baryon Acoustic Oscillations (BAO) of the DESI Collaboration \cite{DESI:2025zgx}, combined with data from the cosmic microwave background (CMB) and Type Ia supernovae (SNIa), have challenged the standard $\Lambda$-Cold Dark Matter ($\Lambda$CDM) model, indicating evidence for a dynamical dark energy component in the Universe. This result is obtained assuming time-dependent equation-of-state (EoS) parameterizations, $w$, defined as the ratio of the dark energy pressure to its density. One such parameterization is the Chevallier-Polarski-Linder (CPL), $w(a) = w_0 + w_a(1-a)$, where $w_0$ and $w_a$ are constants~\cite{Chevallier:2000qy,Linder:2002et}, $a = 1/(1+z)$ is the cosmological scale factor and $z$ is the redshift. In this context, data analysis indicates a conflict at a level $\gtrsim 3\sigma$ with the $\Lambda$CDM model ($w_0 = -1$ and $w_a = 0$).

Parameterizations of dark energy's EoS are not a physical model, as they are not grounded in any underlying physical theory. As a result, for a given combination of $w_0$ and $w_a$, the dark energy density $\rho_{\rm{DE}}$, given in terms of $w$ by $\mathrm{d}\ln\rho_{\rm{DE}} = - 3[1+w(a)]\mathrm{d} \ln a$, may exhibit a phantom behavior at some point in cosmic history, which appears whenever $w(a) < -1$. This concept presents well-known theoretical challenges due to instability issues and violation of energy conditions (see e.g. \cite{Carroll:2003st}).

\begin{figure}[t]
\centering
\includegraphics[width=0.87\columnwidth]{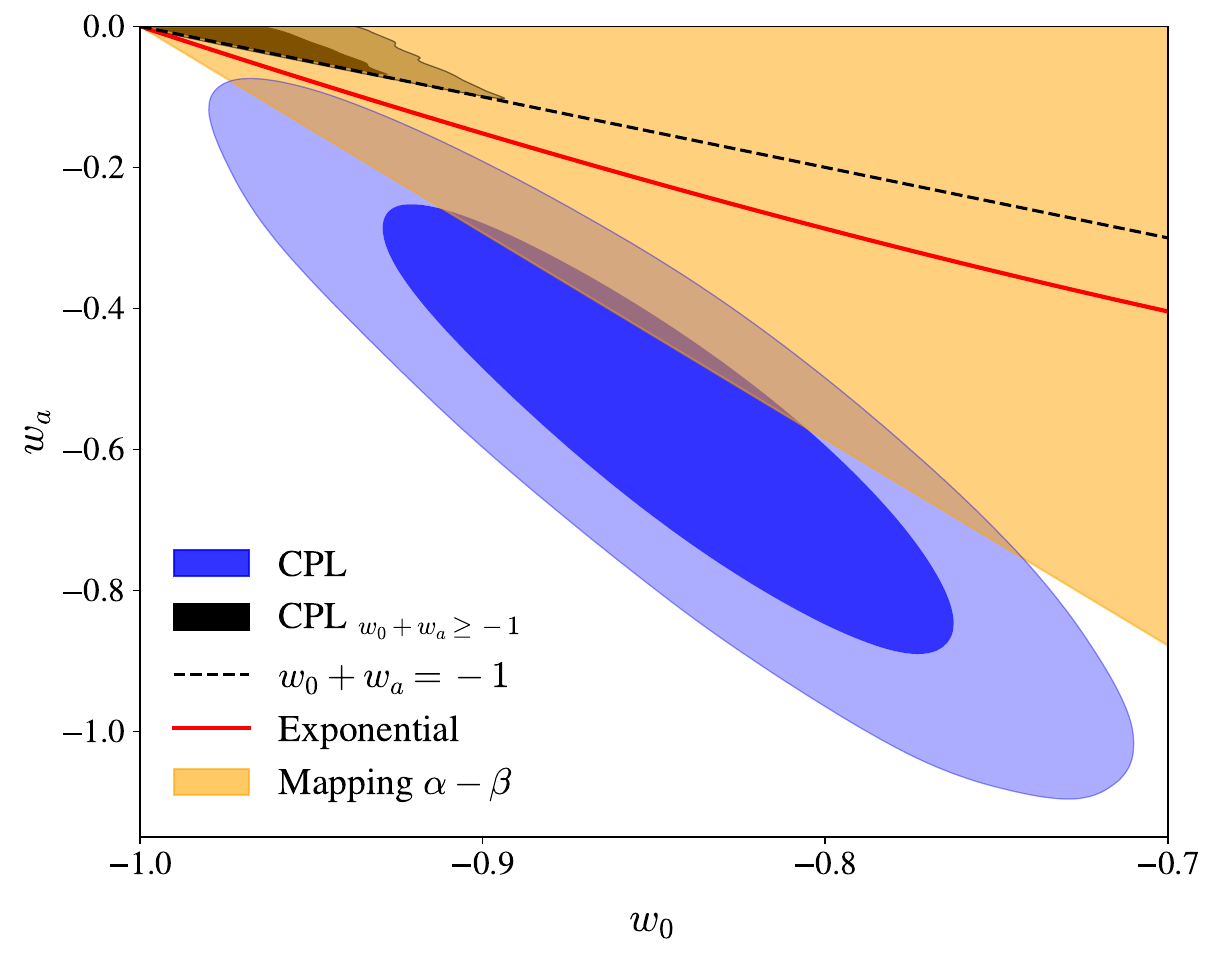}
\caption{A summary of the arguments presented in Sec.~\ref{sec:1}. The figure shows marginalized constraints at 68\% and 95\% C.L. on the $w_0-w_a$ plane from DESI BAO + CMB + Pantheon+ data. The blue contours reproduces the DESI results reported in~\cite{DESI:2025zgx} while the black/dark gray contours on the left upper corner represent the allowed parametric space (68\% and 95\% C.L.) when the non-phantom constraint $w(a) \geq -1$ (which translates to $w_0 + w_a \geq -1$ for CPL) is imposed. As discussed in~\cite{Shlivko:2024llw}, applying this restriction to CPL would exclude a priori families of well-motivated models of thawing quintessence, such as the exponential potential represented by the red line. The orange band represents the region in which the CPL parameterization is mapped into the $\alpha - \beta$ plane with $\alpha, \beta > 0$ -- see Eq.~(\ref{wt}).}
\label{w0wa}
\end{figure}

On the other hand, the background degeneracy between the CPL parameterization and minimally coupled quintessence models, for which $-1 \leq w(a) \leq -1/3$,  maps a given realization of quintessence scenarios into the CPL $w_0-w_a$ plane through the cosmic expansion rate $H(z)$, yielding the same cosmological observables~\cite{Wolf:2023uno,Shlivko:2024llw} - see also \cite{Scherrer:2015tra} for an earlier discussion on this topic. In particular, the $w_0-w_a$ region currently favored by the data predicts $w(a) < -1$ over a period in cosmic evolution. However, it corresponds to the parameter space of classes of  quintessence models, which never crosses the phantom regime by construction. This amounts to saying that rather than concluding that the DESI results point towards a phantom behavior for dark energy, one may interpret the constraints on the $w_0-w_a$ plane as evidence for a dynamical dark energy or, more specifically, for a thawing dark energy component in the Universe. Thus, although simple in form and helpful in searching for deviations from $\Lambda$, CPL is not well-suited to describe the overall behavior of physically-motivated dark energy models, especially of thawing quintessence~\cite{Shlivko:2025fgv} -- a summary of the above arguments is shown in Fig. (\ref{w0wa}). In this context, assessing a possible indication for thawing quintessence from current data outside the lenses of CPL or generic $w_0,w_a$ parameterizations is paramount, as we need physics-based modelings to search for deviations from $\Lambda$ in this precision cosmology era. 

A departure from $\Lambda$ is desired from a theoretical point of view, as a transient cosmic acceleration, characteristic of a thawing behavior, may alleviate some conflicts in reconciling the idea of a dark energy-dominated universe with fundamental physics (see, e.g. ~\cite{Fischler:2001yj,Hellerman:2001yi,Bedroya:2019snp,Cicoli:2023opf} and references therein). For instance, as pointed out long ago by~\cite{Fischler:2001yj,Hellerman:2001yi}, an eternally accelerating universe -- as predicted by the $\Lambda$CDM and some quintessence scenarios -- leads to a cosmological event horizon, which conflicts with string theory predictions as it prevents the description of particle interactions in terms of the standard S-matrix. In other words, the possibility of causally disconnected regions of space implies the absence of asymptotic particle states, which are needed to define transition amplitudes. Furthermore, as recently discussed in \cite{Brandenberger:2025hof}, a bare cosmological constant leads to unitarity problems in the effective field theory framework, which is often formulated as the {trans-Planckian problem} (TPP). Concerning an expanding universe, it arises whenever the comoving Hubble radius $(aH)^{-1}$ decreases, i.e., when there is accelerated expansion. If we live in an eternally accelerating universe, all cosmological scales eventually exit the comoving Hubble radius at a finite time, and the TPP is unavoidable.

Several previous works have explored alternative dynamical dark energy models compatible with recent data from the DESI Collaboration \cite{Ishak:2025cay,Arora:2025msq,Yao:2025wlx,Shah:2025vnt,Qiang:2025cxp,Mishra:2025goj,Sharma:2025qmv,Nojiri:2025low,Gialamas:2025pwv,Ozulker:2025ehg,Cline:2025sbt,An:2025vfz,Linder:2025zxb,Cai:2025mas,Moffat:2025jmx,Bayat:2025xfr,Liu:2025mub,Lin:2025gne,Dinda:2025iaq,Akrami:2025zlb,Kessler:2025kju,Hur:2025lqc,Pan:2025psn,Luu:2025fgw,Ormondroyd:2025iaf}. However, the arguments above underscore the necessity of investigating scenarios that also predict a transient cosmic acceleration. In this sense, we investigate this possibility within the framework of a generalized class of thawing quintessence, which appears to be the most observationally favored behavior of simple quintessence models \cite{DESI:2025fii}. Therefore, this work is organized as follows. In Sec. \ref{thawing section}, we discuss the equation of state of a generalized class of quintessence models, as well as the conditions to achieve transient cosmic acceleration. In Sec. \ref{Datasets&Methodology}, we present the datasets used in the statistical analyses and discuss the adopted methodology. We present our results in Sec. \ref{Results} and our final remarks in Sec. \ref{Final_Remarks}.


\section{Thawing quintessence}\label{thawing section}

The dark energy EoS in standard quintessence scenarios, represented by a single scalar field minimally coupled to gravity with canonical kinetic term, never crosses to the phantom regime. {In the case of thawing quintessence~\cite{Caldwell:2005tm}, it stays frozen with $w \approx -1$ at early times due to Hubble friction, behaving like a cosmological constant. At late times, when the expansion rate is comparable to the field mass, the kinetic term starts to grow and the EoS ``thaws" away to the $w > -1$ region, resulting in $\mathrm{d}w/\mathrm{d}a|_{a =1}>0$, which can then be associated with the $w_a < 0$ region in the CPL plane. However, as opposed to the CPL behavior, the energy density of the quintessence field in this scenario stays roughly constant at early times, followed by a sudden drop to its current value.} An example of model that can explain the late-time cosmic acceleration in this conjecture is the well-known scenario driven by an exponential potential $V(\phi) = V_0 \exp{(-\lambda \phi/M_P)}$ \cite{PhysRevD.37.3406}, where $M_P$ is the Planck mass. Theoretically, this type of potential is motivated in various frameworks, such as supergravity, modified gravity, and superstring theories (see, e.g., \cite{Wetterich:1994bg,Bedroya:2019snp}).

To investigate a broader class of exponential potentials, Ref.~\cite{Carvalho:2006fy} adopted an \textit{ansatz} for the derivative of the scalar field energy density, 
\begin{equation}\label{ansatz}
\frac{1}{\rho_\phi}\frac{\mathrm{d}\rho_\phi}{\mathrm{d}z} = \frac{3\alpha}{(1+z)^{1+\beta}}\;,
\end{equation}
where $\alpha$ is a positive parameter, $\beta$ is nonzero, and the factor of 3 was introduced for mathematical convenience. {The limiting case of $\beta=0$ recovers the exponential scenario proposed in \cite{PhysRevD.37.3406}. Moreover, note that (\ref{ansatz}) essentially reproduces the qualitative features of all slow-roll quintessence models, namely, $\mathrm{d}\rho_\phi/\mathrm{d}z\rightarrow 0$ when $z \gg 1$, rendering an approximately constant energy density at early times, which monotonically decreases with the cosmic expansion, as long as $\alpha > 0$.}

Following the steps developed in \cite{Carvalho:2006fy}, one can obtain for a scalar field-dominated universe $V(\phi) = \Upsilon \exp\left[ -3\alpha\sqrt{\sigma} \left(\phi + \frac{\beta\sqrt{\sigma}}{4}\phi^2\right)\right]$, where $ \Upsilon = V_{0}[1 - \frac{\alpha}{2}(1 + \frac{\beta}{2}\sqrt{\sigma}\phi)^2]$ and $\sigma \equiv 8\pi/3\alpha M_{P}^2$. For all values of $\beta \neq 0$, the above expression represents a class of generalized exponential potentials that admits a broader range of solutions. Among these is a transient accelerating phase, in which the Universe was decelerated in the past, began to accelerate at $z \lesssim 1$, but will return to a decelerating phase in the future.


By combining Eq.~(\ref{ansatz}) with the continuity equation for the scalar field, $\dot\rho_{\phi} + 3 H \rho_\phi[1 + w(a)] = 0$, we may express the dark energy EoS as
\begin{equation}\label{sra}
    w_T(z) = -1 + \alpha (1+z)^{-\beta}\;.
\end{equation}
For $\alpha > 0$, this EoS reproduces with $\sim 1 \%$ or better precision the behavior of various scalar field models as, for instance, those discussed in~\cite{Abreu:2025zng,Shlivko:2025fgv}. Particularly, for $\beta > 0$, it reproduces the behavior of thawing quintessence, in which the EoS derivative with respect to the scale factor is positive at late times. The $\alpha$ parameter controls the present-day value of $w_T$, while $\beta$ dictates how fast it deviates to the $w >-1$ region. From this point onward, our analysis will adopt  (\ref{sra}). The solid curves on the first column of Figure \ref{wt} represent the evolution of  $w_T$ with redshift for the mean values of $\alpha$ and $\beta$ obtained in our statistical analyses (see Table \ref{tab:constraints}). With these parameter values, (\ref{sra}) evolves as $w_T(z \gtrsim 0.5) \approx -1$ followed by a rapid increase to $w_T(z \lesssim 0.5) > -1$. This pattern is approximately what is needed to achieve a fit comparable to the CPL parameterization, as discussed by the DESI Collaboration~\cite{DESI:2025fii}. 

Following the procedure outlined in \cite{Shlivko:2025fgv}, we also obtain the $w_0-w_a$ combination of CPL that better approximates the expansion rate $H(z)$ of a Universe containing a thawing quintessence component described by \eqref{sra}. The orange band in Figure \ref{w0wa} represents the region in which the $\alpha-\beta$ plane is mapped into the $w_0-w_a$ space such that $\alpha,~\beta > 0$. This region fills more of the $w_0-w_a$ plane for increasing values of $\beta$, extending into more negative $w_a$, but it saturates with a subtended inclination $\Delta w_a/\Delta w_0 \approx -2.9$ at $\beta \approx 20$, represented by the lower edge of the orange band. For completeness, we also show how the exponential potential maps into CPL (red curve). As discussed in Sec. \ref{sec:1}, most of the orange band sits below the $w_0 + w_a = -1$ threshold, which supports the arguments for including this region in the $w_0,w_a$ priors of CPL analyses~\cite{Shlivko:2024llw}. Moreover, the exponential potential sits outside the 2$\sigma$ contour of CPL, highlighting the importance of exploring a broader class of quintessence scenarios to obtain a larger overlap with the $w_0-w_a$ constraints.

In addition to being able to successfully account for the present-day cosmic acceleration, the thawing quintessence model results in a Universe that will eventually decelerate in the future. This is a combined effect of the increase in the dark energy EoS followed by an associated drop in its normalized energy density $X_T(z) \equiv \rho_T(z)/\rho_{T,0}$, especially at $z<0$, which makes the dark energy domination come to an end. The strength of this effect, however, depends on the derivatives of the aforementioned quantities, such that a less abrupt evolution of $w_T(z)$ and $X_T(z)$ -- i.e., a lower value of $\beta$ -- will delay the transition to cosmic deceleration. 

To assess a transient accelerating phase, we consider the current age of the Universe $t_{0}$ to be a sensible time scale for the return to a decelerated expansion. That happens whenever the deceleration parameter, defined as $q(z) = -1 + (1 + z) \frac{1}{H(z)} \frac{\mathrm{d}H(z)}{\mathrm{d}z}$, returns to positive values at some time $t < t_0$ in the future. For our class of thawing quintessence, the deceleration parameter reads
\begin{equation}\label{q(z)}
    q(z) = \frac{\Omega_m(1+z)^3 + (1-\Omega_m)X_T(z)[1+3 w_T(z)]}{2[\Omega_m(1+z)^3+(1-\Omega_m)X_T(z)]}\; ,
\end{equation}
where $w_T(z)$ is given by Eq.~\eqref{sra} and $X_T(z) = \exp\{(3\alpha/\beta)[1-(1+z)^{-\beta}]\}$. In Sec. \ref{Results}, we discuss how Eq.~\eqref{q(z)} predicts a transition to a decelerated Universe based on the results of our statistical analyses.


\section{Datasets and Methodology}\label{Datasets&Methodology}

Throughout this work, we use the following CMB, BAO and SNe datasets for our statistical analyses:

\begin{itemize}
    \item Planck: CMB data from Planck 2018 — including low- and high-multipole measurements of temperature and polarization, combined with lensing (plikHM+TTTEEE+lowl+lowE+lensing)~\cite{Planck:2019nip}.
    \item DESI: BAO distance measurements from DESI DR2, which include the bright galaxy sample, luminous red galaxies, emission line galaxies and quasars as tracers. A summary of the measurements is found in Table IV of \cite{DESI:2025fii}.
    
    \item DESY5: Photometrically classified SNe data from the 5 year release of the Dark Energy Survey (DES) collaboration \cite{DES:2024jxu}. The catalog contains 1635 high-redshift ($0.1 < z < 1.3$) and 194 low-redshift ($z<0.1$) SNe. The full catalog, along with supplementary materials, is available through the DES collaboration’s GitHub repository\footnote{\url{https://github.com/des-science/DES-SN5YR}}.
    
    \item Union3: Spectroscopically classified SNe from 24 combined datasets. The resulting dataset is published in a binned format, where the 2087 supernovae are condensed into 22 redshift bins spanning the range $0.05 < z < 0.26$ \cite{Rubin:2023jdq}.
    
    \item Pantheon+: Also a catalog of 1550 spectroscopically classified SNe, compilated from 18 different surveys and covering the redshift range of $0.01 < z < 2.26$ \cite{Scolnic_2022,Brout_2022}. The final dataset and accompanying data products, including the full covariance matrix, are publicly accessible via the Pantheon+ GitHub repository\footnote{\url{https://github.com/PantheonPlusSH0ES/DataRelease/tree/main/Pantheon\%2B_Data}}.
\end{itemize}

For calculating the theoretical predictions, we used a modified version of the publicly available Boltzmann solver CLASS code \cite{lesgourgues2011cosmiclinearanisotropysolving,Diego_Blas_2011}, where we implemented the thawing EoS Eq~\eqref{sra} as an additional dynamical dark energy model.

In order to confront the models with current cosmological data, we perform a Bayesian statistical analysis for $\Lambda$CDM, CPL and thawing quintessence, using the Cobaya \cite{Torrado:2020dgo} \textit{Markov Chain Monte Carlo} (MCMC) sampler. For CPL, we employ the same uniform priors used by the DESI Collaboration, namely $w_0 \in [-3,1]$ and $w_a \in [-3,2]$ \cite{DESI:2025fii}. As for the thawing EoS, we use $\alpha \in [0,1]$, in order to have $w_T(0) \leq 0$, and $\beta \in [0,20]$, to encompass the whole parameter space of the $w_0 - w_a$ mapping - see Figure \ref{w0wa}. We analyze the MCMC chains with GetDist \cite{Lewis:2019xzd} and assess their convergence using the Gelman-Rubin statistic with tolerance of $R -1 < 0.01$ \cite{10.1214/ss/1177011136}. We also perform a statistical model comparison, where we use the MCEvidence package \cite{Heavens:2017afc} in order to calculate the models' Bayesian evidence using the resulting MCMC chains. 

Moreover, the thawing equation of state, as given by Eq.~\eqref{sra}, is a function of the deterministic variable $z$ and of the random variables $\{\alpha,\beta\}$ -- distributions of which are given by the statistical analysis. The same applies to the deceleration parameter, given by Eq.~\eqref{q(z)}, with the additional dependence on the matter density parameter $\Omega_m$. Therefore, we construct the mean evolution and its associated confidence limits of $w(z)$ and $q(z)$ of the thawing quintessence model based on the results of our MCMC chains, for each dataset combination considered. This allows us to better understand the uncertainties associated with the late-time evolution of the EoS, as well as the predictions for a transient accelerating phase in the future.

\section{Results}\label{Results}

\begin{figure*}[t]
\centering
\includegraphics[width=\linewidth]{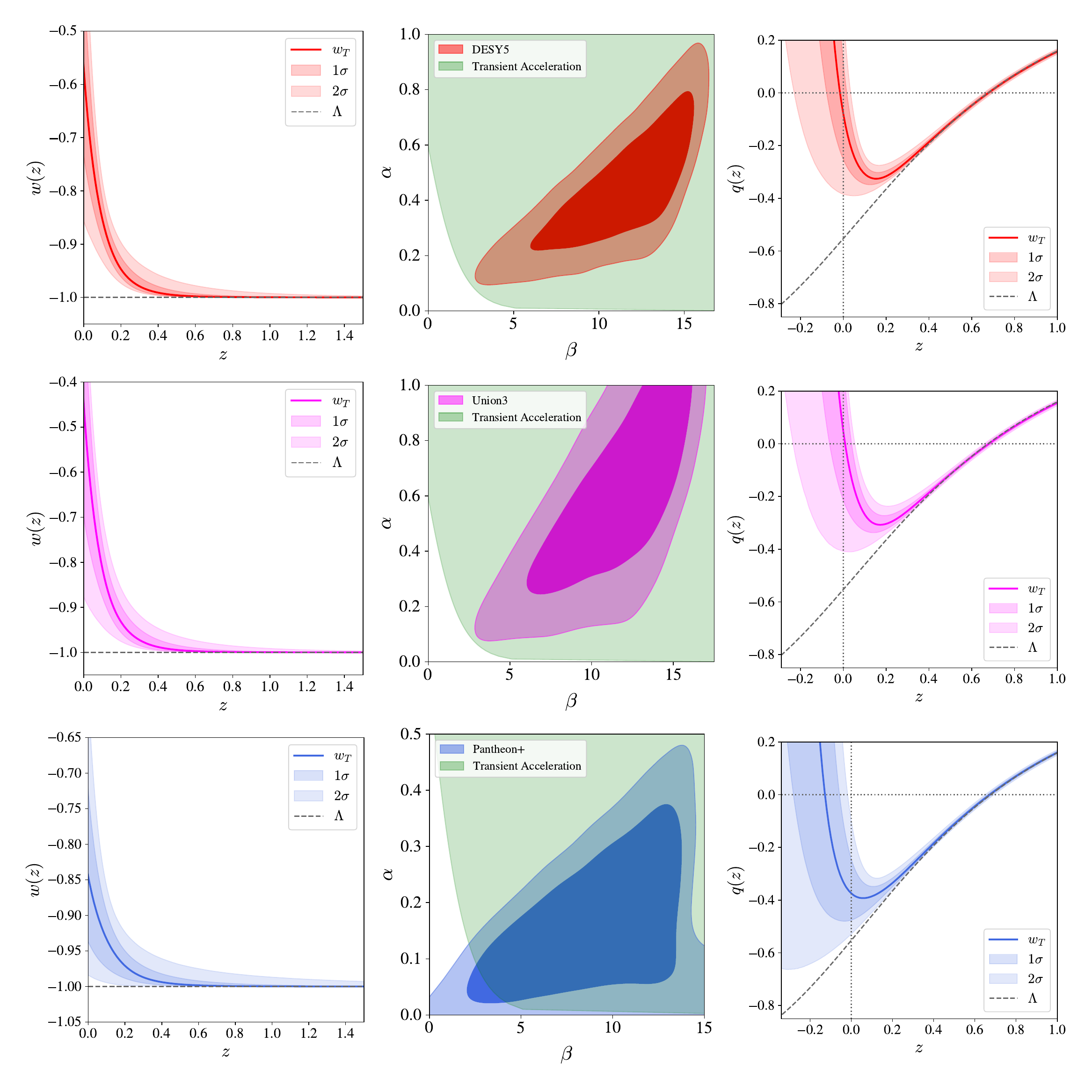}
\caption{The first and last columns show, respectively, the evolution of $w(z)$ and  $q(z)$ as a function of redshift for the mean values of the thawing model discussed in the text, as well as their 1 and 2$~\sigma$ confidence intervals. The middle column shows the marginalized constraints at 68\% and 95\% C.L. on the $\alpha-\beta$ plane. The upper, middle and lower rows refer to each of the SNe samples, namely DESY5 (red), Union3 (magenta) and Pantheon+ (blue), all added to the DESI BAO + CMB data combination. In all cases, the green band delimits the region of transient acceleration. }
\label{wt}
\end{figure*}

\begin{table*}[t]
\centering
\resizebox{\linewidth}{!}{
\begin{tabular}{l c c c c c c c c}
\hline
Dataset & $\Omega_m$ & $h$ & $w_0$ & $w_a$ & $\alpha$ & $\beta$ & $\Delta\chi^2$ & $\Delta\ln B$ \\
\hline
\multicolumn{9}{c}{{$\Lambda$CDM} } \\
DESI+CMB+DESY5 & $0.303 \pm 0.0037$ & $0.682 \pm 0.0029$ & --- & --- & --- & --- & 0.0 & 0.0 \\
DESI+CMB+Union3 & $0.302 \pm 0.0037$ & $0.683 \pm 0.0029$ & --- & --- & --- & --- & 0.0 & 0.0 \\
DESI+CMB+Pantheon+ & $0.302 \pm 0.0036$ & $0.683 \pm 0.0029$ & --- & --- & --- & --- & 0.0 & 0.0 \\
\hline \hline
\multicolumn{9}{c}{{CPL} } \\
DESI+CMB+DESY5 & $0.318\pm 0.0056$  & $0.668\pm0.0056$ & $-0.758\pm0.058$ & $-0.819^{+0.24}_{-0.20}$ & --- & --- & $-23.4$ & $3.45$ \\
DESI+CMB+Union3 & $0.326 \pm 0.0087$ & $0.660 \pm 0.0084$ & $-0.674\pm0.089$ & $-1.05^{+0.31}_{-0.27}$ & --- & --- & $-16.92$ & $1.74$   \\
DESI+CMB+Pantheon+ & $0.310 \pm 0.0057$ & $0.676 \pm 0.0060$ & $-0.843 \pm 0.05$ & $-0.580^{+0.22}_{-0.20}$ & --- & --- & $-7.27$ & $-1.87$ \\
\hline\hline
\multicolumn{9}{c}{{Thawing}} \\
DESI+CMB+DESY5 & $0.322\pm 0.0064$ & $0.661\pm 0.0062$ & --- & --- & $0.45^{+0.15}_{-0.21}$ & $11.3^{+3.8}_{-1.8}$ & $-17.32$ & $1.24$ \\
DESI+CMB+Union3 & $0.327^{+0.010}_{-0.009}$ & $0.656^{+0.0080}_{-0.0110}$ & --- & --- & $0.56\pm 0.24$ & $11.2^{+3.9}_{-2.2}$ & $-10.34$ & $2.40$  \\
DESI+CMB+Pantheon+ & $0.310^{+0.0055}_{-0.0062}$ & $0.674^{+0.0064}_{-0.0057}$ & --- & --- & $0.164^{+0.043}_{-0.16}$ & $9.5^{+4.5}_{-3.4}$ & $-5.38$ & $-0.33$ \\
\hline \hline
\end{tabular}
}
\caption{Results of our statistical analysis for every dataset combination. The last two columns present the values obtained for $\Delta \chi^2$ 
and the Bayes factor, $\Delta \ln B$. 
The above results should be compared to those displayed in Table IV of~\cite{DESI:2025fii}, which includes the CMB ACT data~\cite{ACT:2025fju}. }
\label{tab:constraints}
\end{table*}

Our main results are displayed in Figure \ref{wt}. The middle panel in each row shows the constraints on the $\alpha-\beta$ plane from the combination of DESI DR2 + CMB + designated SN catalog. The tightest constrains were obtained with the DESY5 sample, where we find $\alpha = 0.45^{+0.15}_{-0.21}$ and $\beta = 11.3^{+3.8}_{-1.8}$ at 1$\sigma$ level. Table I shows the results of our statistical analyses for all of the models and datasets. Note that the cosmological constant limit of the thawing quintessence model, namely $\alpha = 0$, is at $\sim2.1\sigma$, $\sim2.3\sigma$ and $\sim1\sigma$ from the mean for DESY5, Union3 and Pantheon+, respectively, indicating no significant deviation from the $\Lambda$CDM model. In contrast, CPL shows a larger discrepancy with $\Lambda$CDM, at  $\sim4.2 \sigma$, $\sim3.7\sigma$ and $\sim2.9\sigma$ level, respectively. This follows the trend discussed in ~\cite{DESI:2025fii,DESI:2025zgx}, where the level of discrepancy of dark energy with $\Lambda$, when considering $w_0w_a$ parameterizations, depends on the choice of the SN sample. Generally, the $\Lambda$CDM model shows more significant discrepancies with the DESY5 and Union3 samples and less deviations with the Pantheon+ compilation. Furthermore, as shown in Fig. \ref{w0wa}, the apparent incompatibility of $\Lambda$CDM with data also depends on the modeling of the dynamical dark energy component, which should be guided by physical considerations~\cite{Abreu:2025zng}.


We also show in Table~\ref{tab:constraints} the results of a Bayesian model selection analysis~\cite{Trotta:2008qt} among the previously discussed scenarios. In this analysis, the strength of evidence is quantified by the Bayes factor, $\ln B_{m_1,m_2} = \ln (\mathcal{Z}{m_1} / \mathcal{Z}{m_2})$, and interpreted using the Jeffreys scale~\cite{Grandis:2016fwl}. There is no preference between the models for $|\ln{B_{m_1,m_2}}| < 1$, whereas for  values of $1<|\ln{B_{m_1,m_2}}| < 2.5$, it shows weak evidence. For $2.5<|\ln{B_{m_1,m_2}}| < 5$ and $|\ln{B_{m_1,m_2}}| > 5$, it gives moderate and strong evidence, respectively. The sign of $\ln B_{m_1,m_2}$ indicates which model is favored: a positive value favors model $m_1$, while a negative value favors model $m_2$. Hence, a negative $\Delta\ln B$ favors $\Lambda$CDM, while a positive value favors the alternative model.

The values of $\Delta\chi^2$ and the Bayes factor for the CPL parameterization and the thawing model are displayed with respect to $\Lambda$CDM. First, the CPL parameterization is moderately favored with DESY5, weakly favored with the Union3 sample and weakly disfavored with the Pantheon+ catalog. These results are consistent with the $\Delta\chi^2$ and Bayes factors reported in Table VI of \cite{DESI:2025zgx} and Table IV of \cite{DESI:2025fii}, respectively, following the same hierarchy of preference over the SNe samples.

Regarding the thawing quintessence scenario, we verify a weak evidence in favor of this model with the DESY5 and Union3 samples, and a weak evidence against it with the Pantheon+ compilation. These results are also somewhat consistent with the values reported in Table III \cite{DESI:2025fii} for their 'algebraic thawing' dark energy parameterization using the deviance information criterion (DIC)~\cite{https://doi.org/10.1111/1467-9868.00353}, although there is a stronger evidence for the more complex model in their case.


Therefore, there is ample parameter space for the $w_T$ EoS consistent with current data, indicating that a broad class of thawing quintessence is competitive with the standard $\Lambda$CDM model while being physically motivated.


In addition, the confidence level in which we can predict a transition to a decelerated Universe also depends on the SNe dataset. For the DESY5 and Union3 catalogs, all of the preferred contours in the $\alpha - \beta$ plane are consistent with transient acceleration -- see the middle panel in the first and second rows in Figure \ref{wt}. This is expected, as the strongest results of DESI in favor of a non-phantom dynamical dark energy require a more abrupt increase of the dark energy EoS at low-$z$~\cite{DESI:2025fii}, which is achieved with higher values of $\beta$ (first panel in the first and second rows). Also, a large $\beta$ is mapped into a more negative $w_a$ in the CPL parameterization - see Fig. \ref{w0wa} and the discussion in Sec. \ref{thawing section} - which is consistent with the results of \cite{DESI:2025zgx} for the same datasets. Hence, these data combination constraints on thawing quintessence seem to be aligned with the parameter space of a faster transition to a decelerated universe, unavoidably yielding transient acceleration at 95\% C. L. On the other hand, the constraints on $\beta$ for the Pantheon+ sample are looser, even allowing small values of this parameter at 1$\sigma$. Therefore, the EoS does not increase fast enough at late-times, which causes cosmic acceleration to persist at the 2$\sigma$ level.


Finally, an important issue worth analyzing concerns the neutrino sector. Although analyses with the CPL parameterization reported in~\cite{DESI:2025ejh} relax the $\Lambda$CDM constraints on $\sum m_\nu$, thus avoiding potential tension with well-established particle physics experiments. They also favor a zero or even a non-physical negative sum of neutrino masses. The results for the thawing EoS (\ref{wt}) within our data choice move in the opposite direction for the constraints, imposing tight bounds on neutrino masses that are comparable to those obtained in the $\Lambda$CDM scenario, while showing the same preference for zero or negative mass. Specifically, our analysis yields $\sum m_\nu < 0.071$~eV at 2$\sigma$ level. For a comprehensive analysis of the thawing model in the presence of neutrinos see~\cite{Rodrigues:2025hso}. 

\section{Final Remarks}\label{Final_Remarks}

Motivated by the recent results from the DESI Collaboration and well-known theoretical arguments for a transient cosmic acceleration, we initially analyzed the evidence for a dynamical dark energy component implicit in the $w_0w_a$ parametric space of the CPL parameterization. Our analysis reinforces the results reported in references ~\cite{Shlivko:2024llw} and \cite{Wolf:2023uno}, which highlight a background degeneracy between general $w_0w_a$ parameterizations and minimally coupled quintessence models. As illustrated in Fig.~\ref{w0wa}, a superficial analysis of this parametric space can be misleading, as it may either exclude families of well-motivated quintessence models or imply a phantom behavior for dark energy \cite{Gialamas:2024lyw}.

Considering a rather general dark energy EoS, capable of reproducing the thawing and freezing behaviors of various scalar field models, we also showed that the current cosmological data can be accommodated with a thawing quintessence model without fine-tuning the parametric space or requiring an exotic phantom behavior for the dark energy component. This model addresses existing discrepancies within the standard model, particularly those related to the neutrino sector, which suggests a zero or even a non-physical negative sum of neutrino mass. Finally, our most robust results, obtained with the DESY5 and Union3 SNe samples, also predict cosmic acceleration as a transient phenomenon at more than 95\% confidence level.

From the standpoint of fundamental physics, such a result offers a way out of the trans-Planckian problem that challenges the ever-accelerated $\Lambda$CDM paradigm by predicting that wavelengths initially smaller than the Planck length will eventually exit the Hubble radius and classicalize, leading to unitarity issues~\cite{Brandenberger:2025hof} (see also \cite{Burgess:2020nec} for a critique to this argument). Also, the cosmological event horizon, i.e., $\Delta = \int{da/a^2H(a)}$, diverges for this transient scalar-field dominated universe, which allows the construction of a conventional S-matrix describing particle interactions within the string theory framework~\cite{Fischler:2001yj,Hellerman:2001yi}.


\section*{Acknowledgements}

RdS and GR are supported by the Coordena\c{c}\~ao de Aperfei\c{c}oamento de Pessoal de N\'ivel Superior (CAPES). JA is supported by CNPq grant No. 307683/2022-2 and Funda\c{c}\~ao de Amparo \`a Pesquisa do Estado do Rio de Janeiro (FAPERJ) grant No. 259610 (2021). We also acknowledge the use of \texttt{class} and \texttt{Cobaya} codes. The development of this work was aided by the National Observatory Data Center (CPDON).

\bibliographystyle{apsrev4-2}
\bibliography{references}

\label{lastpage}

\end{document}